\documentclass[12pt]{iopart}

\usepackage{iopams}  

\usepackage{wrapfig}
\usepackage{graphicx}
\usepackage[font=footnotesize,labelfont=bf]{caption}
\usepackage{subfig}
\usepackage{xcolor}
\usepackage[normalem]{ulem}

\begin{document}

\title[]{Divertor Detachment Characterization in Negative Triangularity Discharges in DIII-D via 2D Edge-Plasma Transport Modeling}

\author{Menglong Zhao$^1$, Filippo Scotti$^1$, Thomas Rognlien$^1$, Marvin Rensink$^1$, Alessandro Marinoni$^2$, Dinh Truong$^1$, Huiqian Wang$^3$, Kathreen Thome$^3$, Carlos Paz-Soldan$^4$}

\address{$^1$Lawrence Livermore National Laboratory, 7000 East Avenue, Livermore, CA 94550 U.S.}
\address{$^2$University of California San Diego, 9500 Gilman Drive, La Jolla, CA 92093 U.S.}
\address{$^3$General Atomics, P.O. Box 85608
San Diego, CA 92186 U.S.}
\address{$^4$Columbia University, 116th and Broadway, New York, NY 10027 U.S.}

\ead{zhao17@llnl.gov}
\vspace{10pt}

\begin{abstract}
Edge fluid modeling of the first divertor-plasma detachment experiments in negative triangularity discharges in the DIII-D tokamak is presented using the 2D multi-fluid edge transport code UEDGE including cross-field particle drifts. Experiments were performed where the lower single-null magnetic equilibrium had a strong negative triangularity ($\delta \approx -0.5$), i.e., where the magnetic X-point is at a larger major radius than the core magnetic axis.  Divertor plasma detachment was induced by increasing the core plasma density in DIII-D via intrinsic gas puffling. Here core density scans are performed with UEDGE to reach a detached plasma and to quantitatively recover the experimental roll-over of the ion saturation current on the outer divertor target-plate. The simulations cover experiments with both signs of the toroidal magnetic field, $B_T$, where the ion magnetic Grad-B drifts are directed into (forward $B_T > 0$) and out of (reverse $B_T <0$) the divertor region.  Consistent with experiments with neutral beam power injection, the negative triangularity simulations reproduce: $~40\%$ higher density is needed to reach detachment onset with forward $B_T$ compared with reverse $B_T$, and the absence of deep detachment is found with reverse $B_T$.  Similarly, comparison between Ohmic discharges in negative and positive triangularity shaping confirms that a substantially higher density is needed to achieve detachment in negative triangularity than in positive triangularity, with negative triangularity requiring an line-average density of at least the Greenwald density or higher. Simulation results suggest that higher densities are needed to reach detachment in negative compared to positive triangularity because these discharges have a shorter midplane-to-target distance along the total magnetic field ${\bf {B}}$, a shorter outer divertor poloidal leg-length ($0.06\,\mathrm{m}$ versus $0.2\,\mathrm{m}$), and reduced radial transport (near-SOL $D_\perp/\chi_\perp = 0.3/0.5$ versus $D_\perp/\chi_\perp = 1.0/1.0$, all in [$\rm{m}^2/\rm{s}$]). 
\end{abstract}

%
%
%
%
%

\section{Introduction}

Plasma power exhaust handling is one of the critical issues for fusion tokamak energy research. In order to achieve fusion reactions, tokamak core plasma is heated to a very high temperature while edge plasma, especially near the divertor target plates, needs to stay below $\sim 10\,\mathrm{eV}$ to reduce peak power deposition on material walls and impurity sputtering rate. One of the proposed operational modes for future power plants is the High-confinement mode (H-mode)~\cite{Hmode} with a steep edge pressure-gradient and detached divertor plasma~\cite{detach} that reduces peak power deposition onto material walls. However, for typical magnetic equilibria used for tokamaks with the poloidal X-point at a smaller major radius than the core magnetic axis (termed positive triangularity), periodic large Edge Localized Modes (ELMs) arise for high input power during H-modes than rapidly transport particles and energy onto open magnetic field lines that rapidly flow to material surfaces ~\cite{ELMs}. ELMs can burn through detached divertor plasmas and deposit high power onto the target plates, damaging material surfaces and enhancing impurity sputtering that can contaminate the core plasma. Even if ELMs can be suppressed, the width of the ELM-free heat flux deposited on target plates is typically much narrower than that in lower-confinement L-mode occurring at lower injection power for positive-triangularity discharges. 

An alternate tokamak operational approach has been suggested by changing the magnetic equilibrium to have negative triangularity (NT) shaping~\cite{Hofmann,Kikuchi,Austin,Coda19,Coda21,Happel,TCV,Kath,NT}.  Here ELMs are typically absent. Further, the core energy confinement time of a NT plasma has been shown to be comparable to that of a H-mode plasma~\cite{Austin,Nelson,Nelson2,Marinoni,Camenen} while the edge/scrape-off layer particle transport appears more typical of L-mode discharges. NT shaping has the attractive features of eliminating large intermittent power fluxes to material walls while having larger radial transport in the edge region to help keep the peak heat-flux on the divertor targets possibly at acceptable levels. Therefore, NT-shaping scenarios have been proposed  for future fusion power plants~\cite{Kikuchi,MANTA}. However, in going to the high power devices, it is likely that operating with a detached divertor plasma will be needed to keep the peak divertor-target heat flux at or below $10~{\rm{MW/m^2}}$. Recent experiments performed on the TCV tokamak found that it is very difficult to obtain detached divertor plasmas with negative-triangularity shaping~\cite{TCV}.  Two stages of plasma detachment are considered here: the first, termed detachment onset, corresponds to the electron temperature ($T_e$) at the target plate reaching $\sim 3-4 \; {\rm eV}$, and the second is deep detachment when $T_e \le 1 {\rm eV}$ and strong plasma recombination occurs resulting in the reduction of the ion-saturation current at the target plate . 
 
A dedicated experimental campaign was recently performed on DIII-D to characterize the operational space in negative triangularity shaping~\cite{Kath,Carlos}. The first detached diverted NT plasma was achieved in this campaign~\cite{Scotti}. The experimental data shows that a continuous degradation of the edge electron temperature $T_e$ pedestal with electron density ramp-up with ion magnetic Grad-B drifts directed into the divertor region (forward $B_T$) even before detachment. With ion magnetic Grad-B drifts directed out of the divertor region (reverse $B_T$), further deep-detachment cannot be attained after reaching detachment onset. For both forward and reverse $B_T$ configurations, DIII-D typically requires a line-average density at a higher fraction of Greenwald density~\cite{Greenwald}, $n_G$,  to reach detachment onset compared to a typical positive triangularity (PT) H-mode discharge for the same input power into the scrape-off-layer (SOL), confirming the difficulty of achieving detached divertor plasmas with a NT shaping.

In order to understand the divertor detachment characteristics mentioned above and the physics behind the difficulty of achieving detachment in the NT shaping, the 2D plasma edge transport code UEDGE is utilized to model several DIII-D NT discharges. These simulations provide an integrated comparison of the behavior of NT divertor plasmas for both signs of $B_T$ and in comparison to PT discharges for similar input parameters. The NT experimental configurations with neutral beam injection where detachment conditions were measured, as well as similar NT and PT Ohmic discharges, are described in Section ~\ref{Exp}. UEDGE modeling of these configurations and comparisons with experimental data is given in Section~\ref{UEDGE}. Discussions and conclusions are made in Sections~\ref{UEDGE} and \ref{summary}.

\section{DIII-D experiment configurations analyzed}\label{Exp}
Two sets of DIII-D experimental configurations and data are used in the comparisons with UEDGE simulations in Sec.~\ref{UEDGE}.  The first set comes from the recent DIII-D NT-campaign where armored tiles were installed on the outer vessel wall to intercept the magnetic separatrix strike points for a diverted NT magnetic equilibrium with neutral beam power injection (NBI)~\cite{Kath}. Here, experiments for both signs of the toroidal B-field are compared.  The second data set comes from NT experiments in lower power Ohmic discharges where the divertor characteristics are compared to a PT discharge with similar parameters providing a comparison of PT and NT configurations.  All NT cases have an approximate negative triangularity of $\delta \approx -0.5$~\cite{Kath,Carlos}.

\subsection{NT cases with neutral-beam power input}\label{NTwNBI}
A strong negative triangularity equilibrium is used for this detachment study with a plasma current $I_p = 0.6~\rm{MA}$ and toroidal magnetic field $B_T = 2~\rm{T}$ (DIII-D discharge \#194288@3000~ms). A density ramp-up was performed here by increasing deuterium gas puffing rate to achieve detached divertor plasma conditions at the outer strike point (OSP). Since most of the power goes into the outboard of the scrape-off layer, the main focus is on outer divertor detachment. The plasma was heated by neutral beam injection $P = 4.3~\rm{MW}$. The core ion $B\times\nabla B$ drift is towards the divertor. As a comparison, a similar discharge (\#194347@3000ms) with the same parameters but with reversing $B_T$, i.e. the core ion $B\times\nabla B$ drift is upward, will be modeled to better understand the effect of B field direction on accessing detachment. More detailed information about these discharges can be found in~\cite{Scotti}. These two discharges are referred to as NT-campaign cases in later discussions.

\subsection{NT and PT cases with only Ohmic power input}\label{ohmic_equil}
In order to compare detachment characteristics of NT and PT, Ohmic discharges with a density ramp in NT shaping ($\delta \approx -0.5$) with ion $B\times\nabla B$ drift driven towards the divertor were performed during the NT-campaign, compared to a discharge with similar setup but a PT shaping from previous experiments. 
In the PT Ohmic discharge, the outer midplane separatrix electron density reaches $n_\mathrm{e,sep} \approx 1.4\times 10^{19}\, \mathrm{m}^{-3}$ when the OSP $J_\mathrm{sat}$ roll-over occurs. However, in the NT ohmic discharge, the line-averaged density is approximately the Greenwald density,  $n_G$, at the end of the discharge, despite the fact that it failed to achieve a detached outer divertor. It can be inferred that it is significantly more challenging to incorporate a detached divertor plasma in the NT shaping than in PT.
 
\section{UEDGE modeling and comparison with auxiliary heated NT discharges for effects of drifts on detachment access}\label{UEDGE}

UEDGE~\cite{UEDGE,UEDGE2} is a 2D transport code that solves the Braginskii equation set~\cite{Brag}  for the plasma density, momentum, and energy assuming classical parallel transport along the magnetic field lines and perpendicular transport including anomalous diffusion and cross-field electric and magnetic drifts. A flux limiter~\cite{limiter} is applied to plasma parallel conductive heat flux and ion viscosity to avoid unphysical high fluxes (from the Spitzer-Harm formula~\cite{SH} for heat flux) when the particle mean-free path is not much less than the local temperature-gradient scale length. Neutral particles are treated as a fluid using a 9-stencil scheme to avoid extreme non-isotropic flows in a non-orthogonal mesh. The main plasma species is deuterium ion $D^+$ and only deuterium atom $D^0$ is considered to be the hydrogenic neutral species in this work. The parallel momentum equation (parallel to $\mathbf{B}$) for $D^0$ is solved due to strong charge-exchange coupling between $D^+$ and $D^0$. Perpendicular transport of $D^0$ is assumed to be diffusive, where the diffusive transport coefficients are obtained based on the assumption that charge-exchange collisions are the dominant processes for $D^0$~\cite{D0diff1,D0diff2,D0diff3}. Carbon impurities are introduced by physical and chemical sputtering from target plates and material walls. Carbon neutral transport is assumed to be diffusive owing to elastic collision, mainly with $D^+$ and $D^0$, with the diffusivity likewise limiter in long mean-free-path regions. Full parallel momentum equations of carbon ions are not solved here. Instead, carbon ion parallel velocities are derived from force-balance analysis of each charge state~\cite{Yu}.

\subsection{NBI NT-discharge characteristics and UEDGE setup}
\begin{figure}
  \centering
  \includegraphics[width=.35\textwidth]{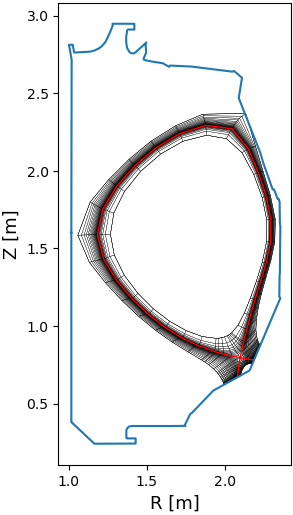}
  \caption{UEDGE grid of the DIII-D negative triangularity discharge (\#194288@3000ms) with neutral beam injection $P = 4.3~\mathrm{MW}$ and separatrix intersecting the vessel wall at the location of the armor tiles.}\label{Fig:3}
\end{figure}

The UEDGE modeling grid, as shown in Fig.~\ref{Fig:3}, for the NT-campaign case is produced based on the reconstructed equilibrium of the shot \#194347@3000ms. The same UEDGE grid is used for the cases with ion $B\times\nabla B$ drift both into and out of the divertor, just with opposite signs of $B_T$ in UEDGE modeling. The core-edge interface is taken at a normalized poloidal flux surface of $\psi_N = 0.9$ and the outermost flux surface is assumed to be at $\psi_N = 1.1$. The radial distance of the outermost boundary from the separatrix at the outer midplane is $\sim 2.7\, \mathrm{cm}$, wide enough to cover $\sim 10\lambda_q$, where $\lambda_q$ is the observed divertor target heat-flux width mapped back to the outer midplane along $\bf{B}$. 

In the NT-campaign shaping with both $B_T$ directions, as mentioned above, plasma is heated by neutral beam $P = 4.3 \, \mathrm{MW}$ and deuterium gas was puffed at the top of the machine to maintain fueling and puff rate was increased to achieve outer divertor detachment. Deuterium density at the core interface is gradually increased to achieve the detached outer divertor plasma. Our previous UEDGE simulations~\cite{Zhao25KSTAR} found that the outer midplane separatrix electron density required to reach an outer divertor detachment onset was insensitive to whether the density ramp-up was achieved by gas puffing at upstream or by increasing the core boundary density.  

\begin{figure}
  \centering
  \includegraphics[width=1.\textwidth]{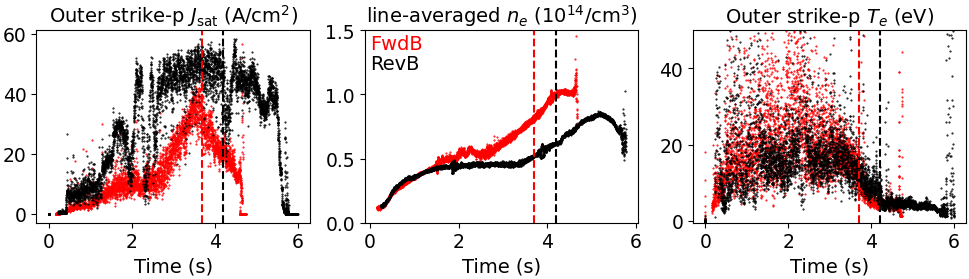}
  \caption{Left: OSP $J_\mathrm{sat}$ as a function of run time for both the forward (red) and reverse (black) $B_T$ cases; Middle: Line-averaged electron density as a function of run time for both the forward (red) and reverse (black) $B_T$ cases; Right: OSP $T_e$ as a function of run time for both the forward (red) and reverse (black) $B_T$ cases. The vertical dashed lines denote the time at which  $J_\mathrm{sat}$ reaches its maximum (typically associated with detachment onset) corresponding to line densities of $1.3\times n_G$ and $1.0\times n_G$ for the forward (red) and reverse (black) $B_T$ cases, respectively, where $n_G$ is the Greenwald density.}\label{Fig:4}
\end{figure}

The experimental OSP ion saturation current density ($J_\mathrm{sat}$) measured by Langmuir probes through the whole discharge time is shown in Fig.~\ref{Fig:4} for forward and reverse $B_T$ cases, with the corresponding core line-averaged electron densities ($n_{e,la}$) shown in the middle panel. There is a clear roll-over of $J_\mathrm{sat}$ with time for the forward $B_T$ case (red). The $J_\mathrm{sat}$ roll-over, commonly regarded as the onset of detachment, occurred around $\sim 3700\, \mathrm{ms}$ when the line-averaged electron density reached to $\sim8\times 10^{19}\, \mathrm{m}^{-3}$. For the reverse $B_T$ case (black), the $n_{e,la}$ rises more slowly in time while the OSP $J_\mathrm{sat}$ initially increases more rapidly, but then saturates. 
Despite the differing temporal trajectories of $n_{e,la}$, the reverse $B_T$ discharge reached the OSP $T_e \sim3-4$~eV ({\it i.e.} detachment onset) at a lower $n_{e,la}$ than the forward $B_T$ discharge (see the right panel of Fig.~\ref{Fig:4}). But unlike the forward $B_T$ case, there is no sustained roll-over of $J_\mathrm{sat}$ for reverse $B_T$ and a further increase in $n_{e,la}$ does not result in deep detachment. 
This so-called 'shallow detachment' for $B_T <0$ is similar to that observed in DIII-D PT discharges~\cite{RevB}. The corresponding line-averaged density is $\sim 6\times 10^{19}\, \mathrm{m}^{-3}$. In comparing NT and PT discharges, it is convenient to use the line-density normalized by the Greenwald density, which scales with plasma current yielding $n_G \approx 6\times 10^{19}\, \mathrm{m}^{-3}$ for both signs of $B_T$ for the NT cases here. Thus, high Greenwald fractions of $\sim 1.3$ and $\sim 1$ are required for detachment onset in forward and reverse $B_T$ configurations, respectively, in this NT shaping.  In comparison, a typical H-mode DIII-D discharge with similar power in PT shaping, begins the  $J_\mathrm{sat}$ roll-over at a Greenwald fraction of $\sim 0.7$~\cite{RevB}. 

\begin{figure}
  \centering
  \includegraphics[width=\textwidth]{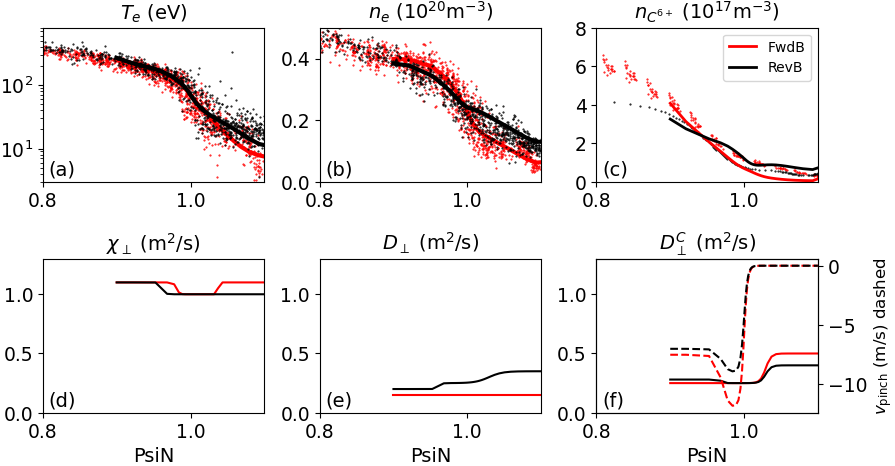}
  \caption{Top row: Thomson scattering data from NT-campaign with NBI (dots) for (a), electron temperature, (b), electron density, and (c),  $n_{C^{6+}}$ density, and UEDGE-computed profiles (solid) at the outer midplane for forward (red) and reverse (black) $B_T$. Second row: UEDGE input for (d), electron/ion thermal diffusivities, (e), deuterium particle diffusivity, and (f), carbon particle diffusivity and pinch velocity (dashed in m/s).}\label{Fig:5}
\end{figure}

%
To determine the anomalous radial plasma transport coefficients to use in UEDGE, the plasma profiles during attached divertor plasmas conditions are used, {\it i.e.},  before $J_\mathrm{sat}$ rollover, corresponding to the line-averaged densities at $6\times 10^{19}\, \mathrm{m}^{-3}$ for forward $B_T$ and $4.3\times 10^{19}\, \mathrm{m}^{-3}$ for reverse $B_T$ (see Fig.~\ref{Fig:4}).  The corresponding near-midplane experimental Thomson scattering electron density and temperature profiles are shown in the first row of Fig.~\ref{Fig:5} as dots.  The second row shows the inferred UEDGE radial diffusivity profiles of (equal) ion and electron thermal diffusivities, $\chi_\perp$, and particle diffusivity, $D_\perp$ that yield comparable UEDGE profiles in the second row. A small adjustment is also made for the input power through the core-edge interface. The resulting power used is $4\, \mathrm{MW}$, close to the experimental power input of $4.3~\rm{MW}$. Similarly, the anomalous particle diffusivity coefficients of carbon ions , $D_\perp^\mathrm{C}$, together with a pinch velocity $v_\mathrm{pinch}^\mathrm{C}$ are obtained by fitting the experimental radial density profile of $C^\mathrm{6+}$. All carbon ions are assumed to have these same transport coefficients. The anomalous diffusivity coefficient and pinch velocity used are assumed to be poloidally constant between the inner and outer divertor entrance. In the divertor regions below the magnetic X-point, a constant value is used for electrons and all ion species: $D_\perp = D_\perp^\mathrm{C} = 1.0\,\mathrm{m}^2/\mathrm{s}$, $\chi_\perp^\mathrm{e} = \chi_\perp^\mathrm{i} = 1.0\,\mathrm{m}^2/\mathrm{s}$. The detachment characteristics are insensitive to moderate variation of $D_\perp$ and $\chi_\perp$ in the divertor region in the simulations because the perpendicular transport arising from the $\rm{E\times B/B^2}$ drift velocity dominates radial transport there.

\begin{figure}
 \centering
 \includegraphics[width = 0.66\textwidth]{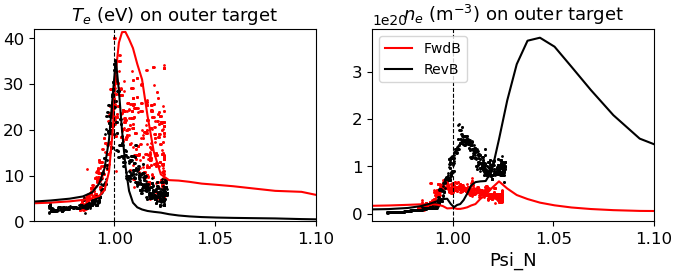}
 \caption{Electron temperature (left) and density (right) along the outer target plate from both the experiment (dots) and UEDGE simulation (solid) for both the forward (red) and reverse (black) $B_T$ configurations}\label{Fig:6}
\end{figure}


\subsection{Radial plasma profiles and detachment onset: experiments vs simulations}
With the radially varying profiles of $D_\perp$ and  $\chi_\perp^\mathrm{e,i}$ [see Fig.~\ref{Fig:5}(d,e)], a good match between the simulated and experimental radial profiles of  electron density and temperature near the outer midplane is achieved as shown by the solid lines in Fig.~\ref{Fig:5}(a,b). While comparable upstream density profiles of $C^\mathrm{6+}$ between the experiment and modeling inside the separatrix can be obtained using radially varying $D_\perp^\mathrm{C}$ and $v_\mathrm{pinch}^\mathrm{C}$, there is more of a mismatch between the profiles of $C^\mathrm{6+}$ in the outer SOL region that cannot be easily removed by adjusting $D_\perp^\mathrm{C}$ and $v_\mathrm{pinch}^\mathrm{C}$. The majority of the carbon in this region resides in lower charge states.  Nonetheless, the physics behind the mismatch for $C^\mathrm{6+}$ is not fully understood and is beyond the scope of this paper. The modeling and experiments are comparable on the electron temperature along the outer target plate while the electron density peak from modeling is shifted towards the far-SOL by $\sim 2-3\,\mathrm{cm}$, as shown in Fig.~\ref{Fig:6}. This discrepancy seems to be a general issue at low collisionality that has not been fully understood~\cite{PSI}, though it is known to be sensitive to the treatment of plasma recycling on the radial simulation boundary near the outer target plate. The radial ${\bf E \times B}$ flow adjacent to the target plate was identified as the mechanism that drives plasma away from OSP into far-SOL for reverse $B_T$ similar to~\cite{RevB}.  In contrast, with forward $B_T$, the plasma at the OSP is driven into the inner divertor through the private-flux region, but the reduced density is not replenished by the radial ${\bf E \times B}$ flow from flux tubes at larger major radii. The overall effects lead to an radially outward shift of the peak density on the outer target plate. In this simulation, $D_\perp$, $D_\perp^\mathrm{C}$, $\chi_\perp^\mathrm{e,i}$, and $v_\mathrm{pinch}^\mathrm{C}$ profiles were fixed,  assuming that variation of these coefficients is negligible, while increasing upstream collisionality to achieve detachment in UEDGE. Due to this density shift, we will use the roll-over of the integrated ion saturation current $I_\mathrm{sat}$ on the outer target plate as an indicator of detachment onset.

\begin{figure}
  \centering
  \includegraphics[width=.66\textwidth]{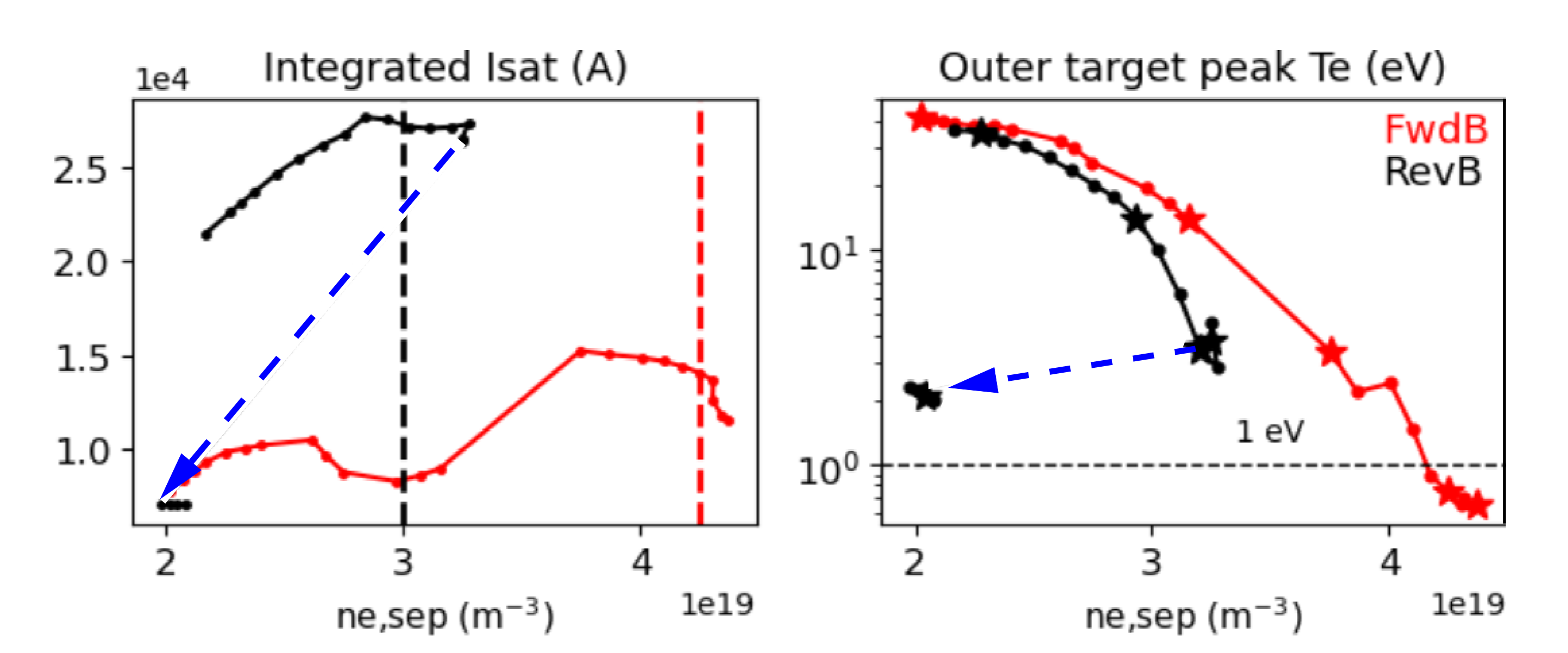}
  \caption{Left: Integrated ion saturation currents ($I_\mathrm{sat}$) with the increase of outer midplane separatrix electron density for both the forward (red) and reverse (black) $B_T$ configurations from UEDGE simulations. The two vertical dashed lines denote $n_\mathrm{e,sep}$ at the roll-over of OSP $J_\mathrm{sat}$ observed in the experiments for the forward (red) and reverse (black) $B_T$ configurations. Right: Peak electron temperature on the outer target plate as the increase of outer midplane separatrix electron density for both the forward (red) and reverse (black) $B_T$ configurations from UEDGE.}\label{Fig:7}
\end{figure}

The outer divertor detachment is achieved in UEDGE at a similar outer midplane separatrix electron density $n_\mathrm{e,sep}$, compared to the experimental data for both forward and reverse $B_T$ cases, shown in Fig.~\ref{Fig:7}. However, a detachment bifurcation, i.e. the peak $T_e$ on the outer target plate drops rapidly from $\sim 10\,\mathrm{eV}$ to $\sim 1\,\mathrm{eV}$~\cite{bifur1,bifur2,bifur3,bifur4,bifur5,bifur6}, occurs for the forward $B_T$ case in the simulations although there is no evidence for the existence of an abrupt bifurcation in the experiment. As in the experiments, the simulations find that a higher $n_\mathrm{e,sep}$ is required to reach detachment onset for the forward $B_T$ than for reverse $B_T$. The simulations show that a higher upstream density $n_\mathrm{e,sep}$ is needed for detachment in the outer divertor for forward $B_T$ because then the plasma is driven away from the outer divertor through the private flux region into the inner divertor by $E\times B$ flows, similar to regular PT shaping~\cite{RevB,Rognlien,Boedo,Pitts,Rozhansky,Chankin2015,Wensing,LHD,EAST}. 
\subsection{Evolution of detahcment in NT density ramp}\label{sec:2D}
\begin{figure}
  \centering
  \includegraphics[width=\textwidth]{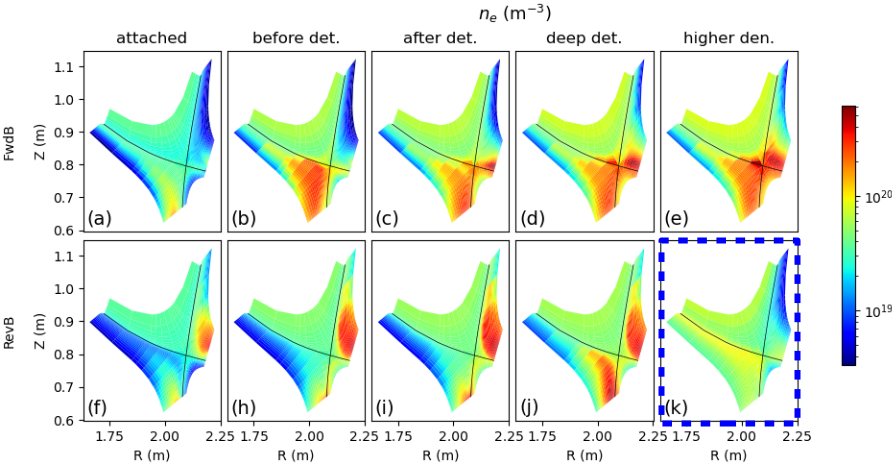}
  \caption{2D profiles of electron density for both the forward ((a), (b), (c), (d) and (e)) and reverse ((f), (h), (i), (j) and (k)) $B_T$ configurations with increasing densities. The five chosen cases: 'attached', 'before det.', 'after det.', 'deep det.', 'higher den.', are denoted by stars in Fig.~\ref{Fig:7}. Details about the five cases both for the forward and reverse $B_T$ configurations can be found in the text. The case (k) with the blue dashed rectangular box denotes the case after thermal collapse.}\label{Fig:9n}
\end{figure}

\begin{figure}
  \centering
  \includegraphics[width=\textwidth]{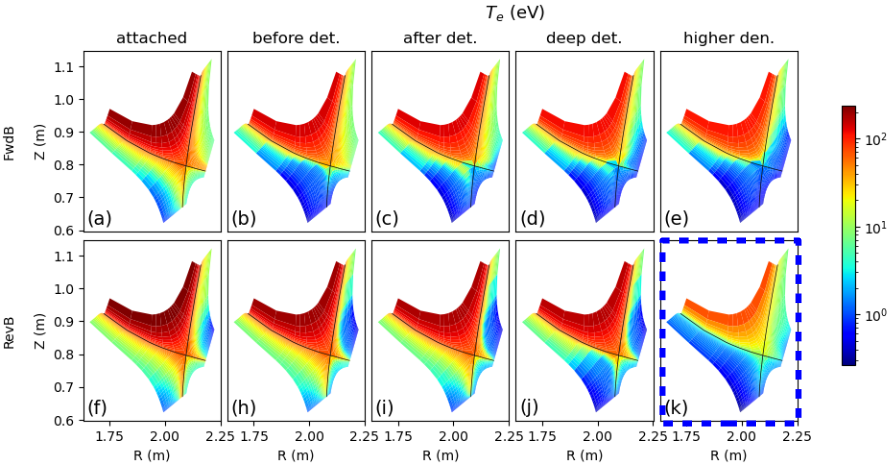}
  \caption{2D profiles of electron temperature for both the forward ((a), (b), (c), (d) and (e)) and rverse ((f), (h), (i), (j) and (k)) $B_T$ configurations with increasing densities. The five chosen cases 'attached', 'before det.', 'after det.', 'deep det.', 'higher den.' are the same as in Fig.~\ref{Fig:9n}. The case (k) with the blue dashed rectangular box denotes the case after thermal collapse.}\label{Fig:9T}
\end{figure}

To elucidate the distinct behaviors observed in forward and reverse $B_T$ configurations---namely, the gradual performance degradation in forward $B_T$ and the absence of deep detachment in reverse $B_T$---we analyze the evolution of detachment in both cases. Figure~\ref{Fig:9n} shows the 2D profiles of electron density for forward $B_T$ [top row: (a, b, c, d, e)] and for reverse $B_T$ [second row: (f, h, i, j, k)]. The first through fourth columns represent a monotonic increase in upstream density, corresponding respectively to attached (a,f), before detachment onset (b,h), after detachment onset (c,i), and deeply detached outer divertor (d,j) conditions. The fifth column (e,k) corresponds to a case with an upstream density even higher than that in the fourth column. The five cases correspond to the five steady states indicated by star symbols in Fig.~\ref{Fig:7}. The corresponding $T_e$ contours are shown in Fig.~\ref{Fig:9T}. For the forward $B_T$ configuration, the five cases are chosen to cover various divertor plasma states, from very attached where the outer target peak $T_e$ is $\sim 30-40\,\mathrm{eV}$, through a high recycling state with peak $T_e$ $\sim 10\,\mathrm{eV}$ after which no solution is found until $n_{e,sep} \approx 3.8~\times 10^{19}{\rm{m^{-3}}}$ where the plasma approaches detachment near $T_e$ $\sim 3\,\mathrm{eV}$, and then enters deep detachment where $T_e$  stays below $1\,\mathrm{eV}$ (see Figs.~\ref{Fig:9T}). The lack of a steady-state solution between $n_{e,sep} = 3.2-3.8 \times 10^{19}{\rm{m^{-3}}}$ may be associated with a bifurcation in the numerical solution, but that is not explored here. For the last two solutions in upper row (d,e) in Figs.~\ref{Fig:9n} and \ref{Fig:9T}, $T_e$ in both inner and outer divertor are $\lesssim 1\,\mathrm{eV}$ (see Fig.~\ref{Fig:9T}). Turning to the reverse $B_T$ configuration, the first three cases are chosen so to have comparable outer target peak $T_e$ to the cases in forward $B_T$. Because there is no clear roll-over of $I_\mathrm{sat}$ in reverse $B_T$, consistent with the experimental observation, and the OSP $T_e$ (the same as peak $T_e$) decreases only to $\sim 2-3\,\mathrm{eV}$ as upstream density increases and no further decrease can be attained until a wide-spread thermal 'collapse' occurs (as discussed in the following section), the fourth (j) and fifth (k) cases in reverse $B_T$ correspond to  before and after the thermal collapse where $n_e$, $T_e$, and radiation all change substantially, as discussed in Sec.~\ref{collapse}.
\begin{figure}
  \centering
  \includegraphics[width=\textwidth]{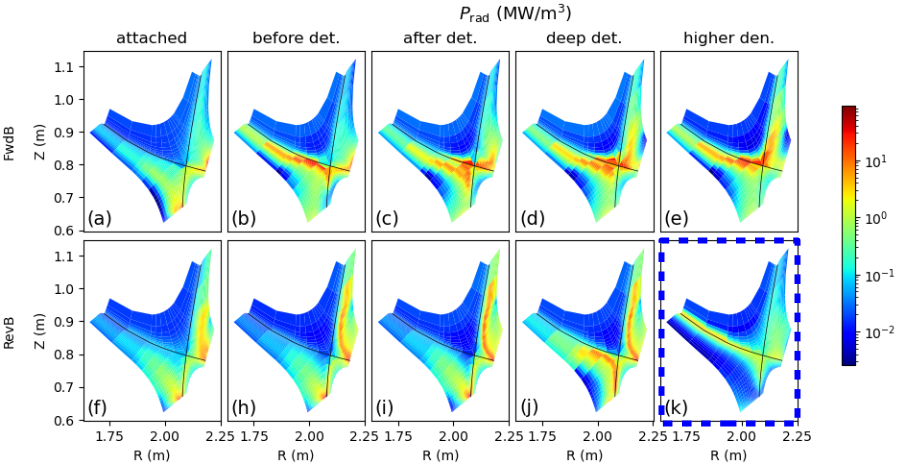}
  \caption{2D profiles of radiation for both the forward ((a), (b), (c), (d) and (e)) and reverse ((f), (h), (i), (j) and (k)) $B_T$ configurations with increasing densities. The five chosen cases 'attached', 'before det.', 'after det.', 'deep det.', 'higher den.' are the same as in Fig.~\ref{Fig:9n}. The case (k) with the blue dashed rectangular box denotes the case after thermal collapse.}\label{Fig:rad2D}
\end{figure}

Figures~\ref{Fig:9n}(a,b) and (f,h) show that the electron density in the outer divertor with forward $B_T$ is considerably lower than that with reverse $B_T$ for similar upstream collisionalities before reaching detached outer divertor. The evolution of electron density in both divertors is very different between the two configurations, mainly due to different directions of the $E\times B$ drift. Focusing first on forward $B_T$, Fig.~\ref{Fig:9n} (a) and (b) show that $n_e$ in the inner divertor increases rapidly before detachment due to particles moved from the outer to the inner divertor through the private flux region by $E\times B$ drifts. Correspondingly the inner divertor becomes cooler rapidly as shown in Fig.~\ref{Fig:9T} (a) and (b). When the high density region on the high-field-side (HFS) expands across the separatrix, there is a sharp increase in $n_e$ and decrease in $T_e$ on the outer target near the strike point, with peak $n_e$ increasing from $\sim 1\times 10^{20}\, \mathrm{m}^{-3}$ to $\sim 5\times 10^{20}\, \mathrm{m}^{-3}$ and peak $T_e$ dropping from $14\,\mathrm{eV}$ to $3\,\mathrm{eV}$, with the final state coinciding to the onset of detachment [see Figs.~\ref{Fig:9n} (c) and~\ref{Fig:9T} (c)]. As $n_{e,sep}$ continues to increase, $n_e$ in the outer divertor keeps increasing [see Fig.~\ref{Fig:9n} (d) and (e)] and $T_e$ drops further with the region of $\lesssim 1\,\mathrm{eV}$ extending towards the X-point [Fig.~\ref{Fig:9T} (d) and (e)], while $T_e$ remains low on the HFS. Here deep detachment is sustained at both divertor targets. 

The evolution of $n_e$ and $T_e$ in the inner and outer divertors behave nearly opposite in reverse $B_T$. $n_e$ in the outer divertor is pushed towards the far SOL and initially increases rapidly as the core-edge boundary is increased [Fig.~\ref{Fig:9n} (f), (h) and (i)], until the outer target peak $T_e$ (or OSP $T_e$) reaches $\sim 2-3\,\mathrm{eV}$. No further decrease of OSP $T_e$ is attained as the upstream density increases, as shown in Fig.~\ref{Fig:7} and Fig.~\ref{Fig:9T} (i) and (j), similar to the analysis of a PT discharge in reverse $B_T$~\cite{RevB}. However, $n_e$ in the inner divertor then begins to accumulate rapidly [Fig.~\ref{Fig:9n} (i) and (j)] because the $E\times B$ driven particle flux flowing through the private-flux region from the inner to outer divertor regions is reduced because of the decreasing $T_e$, thus lower electrostatic potential, $\phi$, in divertor regions.  The radiation front begins moving upstream along the separatrix [Fig.~\ref{Fig:rad2D} (i) and (j)]. Nevertheless, unlike a stable deep detached divertor that can be maintained with the forward $B_T$ configuration, a more global thermal collapse abruptly appears for reverse $B_T$ with a further increase of the core density boundary condition, as shown in Figs.~\ref{Fig:7}, \ref{Fig:9n}(k), \ref{Fig:9T}(k) and \ref{Fig:rad2D}(k).  Some details of this thermal collapse follow in the next subsection.
\subsection{Shallow detachment in reverse $B_T$ and impact on core-edge plasma including confinement}\label{collapse}
\begin{figure}
  \centering
  \includegraphics[width=.9\textwidth]{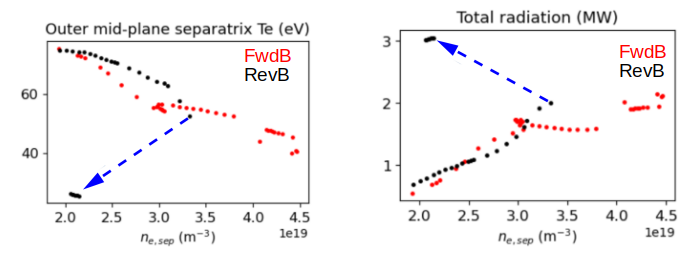}
  \caption{Outer midplane separatrix electron temperature $T_\mathrm{e,sep}$ (left) and total radiation (right) as the increase of outer midplane separatrix electron density for both the forward (red) and reverse (black) $B_T$ configurations. The arrows indicate a simultaneous bifurcation-like change of $T_\mathrm{e,sep}$ and total radiation when a certain upstream collisionality is reached.}\label{Fig:10}
\end{figure}

After reaching detachment onset further deep detachment cannot be attained experimentally with reverse $B_T$, which is reproduced in the UEDGE simulations shown in Fig.~\ref{Fig:7}. As the upstream collisionality increases, the integrated ion saturation current $I_\mathrm{sat}$ first increases, and then saturates until an abrupt drop by a factor of 4. The bifurcation-like drop of $I_\mathrm{sat}$ is not a typical characteristic of a deep detachment. Figure~\ref{Fig:10} shows the evolution of the outer midplane separatrix electron temperature ($T_\mathrm{e,sep}$) and total radiation with increasing core-edge interface density. With an incremental increase in $n_{core}$, a bifurcation-like $50\%$ reduction in $n_\mathrm{e,sep}$ occurs while  $T_\mathrm{e,sep}$ drops from $\sim 50\, \mathrm{eV}$ to $\sim 20\,\mathrm{eV}$. Similarly, $I_\mathrm{sat}$ drops as noted just above. The total radiation increases dramatically at the same time, and the radiation front moves along the separatrix toward the midplane on the HFS (see the bottom panel in the fifth column of Fig.~\ref{Fig:rad2D}), indicating that a thermal collapse occurs. As discussed in Sec.~\ref{sec:2D}, OSP $T_e$ is kept above $2\,\mathrm{eV}$ (see Fig.~\ref{Fig:9T}) due to density shifting towards far SOL (see Fig.~\ref{Fig:9n}) with the increase of upstream collisionality until reaching a thermal collapse. The simulations suggest there is no steady state solution for a deeply detached outer divertor in the reverse $B_T$ NT configuration. This overall picture  is very similar to that from simulations of a PT-shaped discharge~\cite{RevB}.

\begin{figure}
  \centering
  \includegraphics[width=0.9\textwidth]{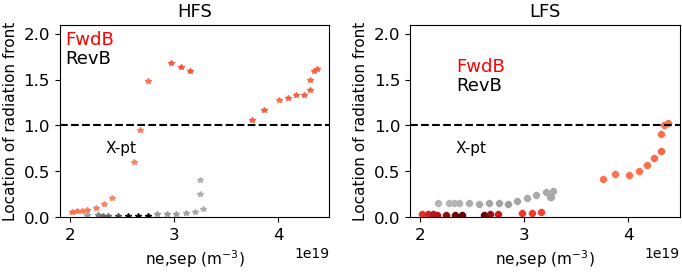}
  \caption{The normalized distances between the peak radiation locations and target plates on both HFS (left) and LFS (low-field side, right) as a function of $n_\mathrm{e,sep}$ are shown as stars and circles, respectively, with the forward (red) and reverse (black) $B_T$ configurations from UEDGE simulations. The normalization lengths are inner poloidal leg length for HFS radiation front and outer poloidal leg length for LFS radiation front, respectively. The location of the X-point is corresponding to unity (dashed line). The two-tone brightness of the dots denotes relative strength of the radiation.}\label{Fig:radfront}
\end{figure}

\begin{figure}
  \centering
  \includegraphics[width=0.5\textwidth]{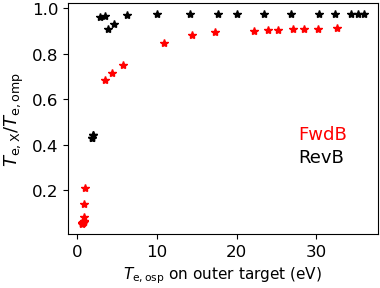}
  \caption{The ratio of lowest $T_e$ inside the last-closed-flux-surface (LCFS) above the X-point (dot) to outer min-plane separatrix electron temperature $T_\mathrm{e,omp}$ (star) as functions of OSP $T_e$ with the forward (red) and reverse (black) $B_T$ configurations.}\label{Fig:Tx}
\end{figure}
In the experiment, confinement degradation was observed for detached conditions beyond the onset of detachment with both the forward and reverse $B_T$ configurations~\cite{Scotti}. Especially in the forward $B_T$, the confinement degradation started before the onset of detachment with a continuous degradation of the edge $T_e$ pedestal. The UEDGE simulations above show that $n_e$ in the inner divertor increases rapidly as upstream density increases in forward $B_T$, and the high density region extends above the X-point in the HFS SOL even before the detachment of the outer divertor, creating a HFS highly radiating, low $T_e$ region (see Figs.~\ref{Fig:9n} (b),~\ref{Fig:9T} (b) and ~\ref{Fig:rad2D} (b)). The HFS radiation front moves above the X-point in the SOL even when the outer divertor is attached, shown in Fig.~\ref{Fig:radfront}, cooling the plasmas in the vicinity of the X-point, decoupling the electron temperature above the X-point inside the separatrix $T_\mathrm{e,X}$ and outer midplane electron temperature $T_\mathrm{e,omp}$ shown in Fig.~\ref{Fig:Tx}, before reaching outer divertor detachment. At detachment onset, the HFS radiation front moves across the separatrix, stabilizing right above the X-point (see Figs.~\ref{Fig:rad2D} (c) and~\ref{Fig:radfront}), further reducing $T_\mathrm{e,X}$ sharply to $\lesssim 10\,\mathrm{eV}$ ($T_\mathrm{e,X}/T_\mathrm{e,omp} \lesssim 0.2$) as upstream density increases. This is qualitatively consistent with experimental observations. However, in reverse $B_T$, the decoupling of $T_\mathrm{e,X}$ and $T_\mathrm{e,omp}$ and a clear reduction of both only occur after the thermal collapse in the UEDGE simulations, which seems to be different from the experimental data.

\section{Understanding detachment in comparable NT and PT Ohmic discharges}
\begin{figure}
  \begin{minipage}{0.4\textwidth}
  \centering
  \includegraphics[width=\textwidth]{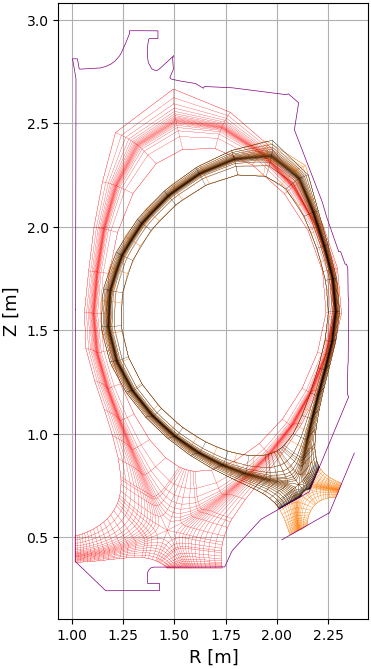}
  \caption{UEDGE grids for the experimental  NT (black) and PT (red) shapings, and an NT shaping with artificially extended legs (orange) to assess effect of divertor-leg length.}\label{NTPT_grids}
  \end{minipage}
\end{figure}

The requirement of $n/n_G > 1$ for the roll-over of OSP $J_\mathrm{sat}$ in the NT-campaign shaping with NBI injection reveals the difficulty of achieving a detached outer divertor in NT discharges, which is further confirmed by the results of the aforementioned Ohmic discharges in both NT and PT shaping. In order to better understand the physics behind this, UEDGE is used to simulate two similar Ohmic discharges, one with NT shaping and the other with PT shaping.  The UEDGE flux-surface grids for both the NT (\#193831@1900s) and PT (\#175194@2800s) magnetic equilibria are shown in Fig.~\ref{NTPT_grids}, The UEDGE setup for these discharges is similar to the NBI discharges already discussed, but with a lower power input of $P = 1.6\,\mathrm{MW}$ at the core-edge interface and different radial transport coefficients shown below. Here, the core ion gradient-B drift is always into the X-point, {\it i.e.}, $B_T >0$. 

\begin{figure}
  \centering
  \includegraphics[width=0.7\textwidth]{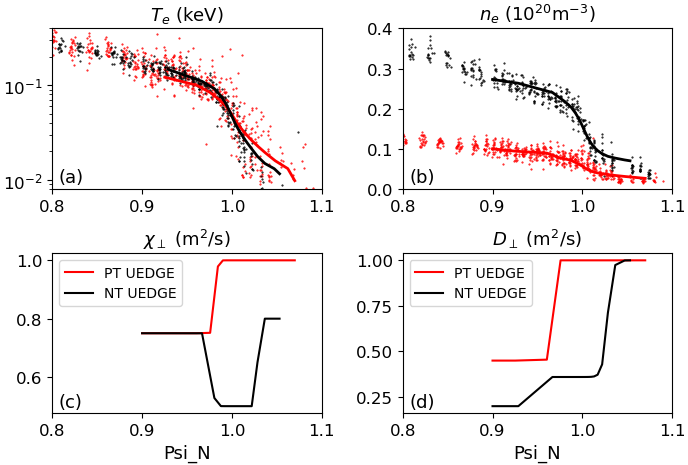}
  \caption{Electron temperatures (a) and densities (b) from experiments (dots) and UEDGE modeling (solid) along the outer midplane for both the NT (black) and PT (red) cases. The UEDGE profiles are obtained using the profiles of the radially varying thermal (c) and particle (d) diffusivities.}\label{Fig:12a}
\end{figure}

Similar to the NT-campaign NBI cases, two attached divertor-plasma states are simulated and radial plasma transport coefficients are deduced by fitting UEDGE profiles of outer midplane $n_e$ and $T_e$ to experimental profiles derived from Thomson Scattering (see Fig.~\ref{Fig:12a}). The resulting $D_\perp$ and $\chi_\perp$ profiles for each shaping shown in Fig.~\ref{Fig:12a}. The $D_\perp$ and $\chi_\perp$ derived for the NT case are $\sim 65\%$ and $\sim 50\%$ lower than those in the PT case near the separatrix between $\psi_N = 0.97$ and $\psi_N = 1.02$. 

\begin{figure}
  \centering
  \includegraphics[width=.7\textwidth]{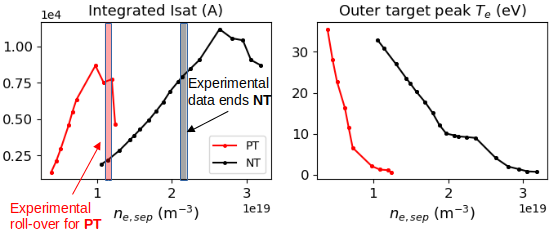}
  \caption{Integrated ion saturation currents $I_\mathrm{sat}$ (left) and peak $T_e$ on the outer target plate (right) as increasing outer midplane separatrix electron density for both the NT (black) and PT (red) cases.}\label{Fig:13}
\end{figure}

The set of UEDGE steady-state solutions for increasing $n_{e,sep}$ are quantitatively consistent with the experimental data with regard to the rollover of the $J_\mathrm{sat}$ signal, which is used to identify detachment onset in Fig.~\ref{Fig:13},  {\it i.e.}, the NT configuration requires a substantially higher $n_{e,sep}$ than the PT configuration. The $n_\mathrm{e,sep}$ at the roll-over point of $J_\mathrm{sat}$ is close to the experimental value of $\approx 1.4 \times 10^{19}{\rm m^{-3}}$ in the PT shaping.  For the NT shaping, the experiment only reaches $n_{e,sep} \approx 2.2 \times 10^{19} {\rm m^{-3}}$ without showing a rollover of $J_{sat}$, while the simulations go to higher $n_{e,sep}$ and show a roll-over at $n_{e,sep} \approx 2.6 \times 10^{19} {\rm m^{-3}}$ . There are three differences between the NT and PT configurations that could contribute to the different detachment characteristics of the NT and PT shaping:
\begin{enumerate}
\item Geometry effect: The distance along $\bf{B}$ between the outer midplane and the outer target (the connection length) for the NT case is $\sim 1/3$ of that in the PT case. Also, the divertor volume is much smaller in the NT case owing to the shorter divertor leg-length (see Table 1).

\item Radial transport effect: $D_\perp$ and $\chi_\perp$ of the NT plasma are lower by $\sim 65\%$ and $\sim 50\%$, respectively.

\item Smaller $|B|$ in the divertor: The divertor of the NT shaping is at a larger radius. On average, the magnetic field in the divertor of the NT case is lower than that of the PT, which could result in higher $\bf{E\times B}/B^2$ flows.
\end{enumerate}

\begin{figure}
  \centering
  \includegraphics[width=.8\textwidth]{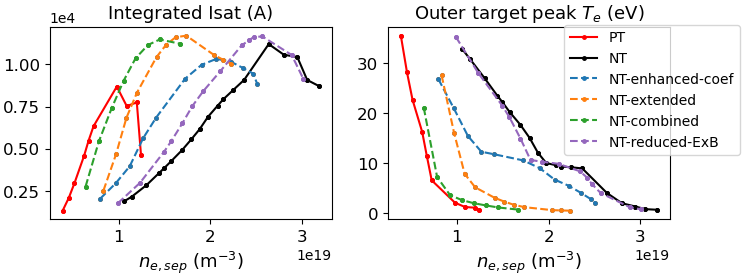}
  \caption{Integrated ion saturation currents $I_\mathrm{sat}$ (left) and peak $T_e$ on the outer target plate (right) as increasing outer midplane separatrix electron density for the NT (black), PT (red) cases, NT with extended legs (orange), NT with enhanced transport coefficients (blue), NT with both effects (green) and NT with reduced ExB drifts (purple).}\label{Fig:14}
\end{figure}

When keeping the strong negative triangularity ($\delta \approx -0.5$), it is difficult to change the connection length for the NT shaping in the experiment, which is mainly determined by the plasma current and the location of the X-point. However, in UEDGE the divertor target plates can be moved freely without affecting the equilibrium. Therefore the effect of a larger divertor volume is assessed by simulations where both inner and outer plates are moved to allow longer divertor legs in the NT shaping as shown by the orange mesh in Fig.~\ref{NTPT_grids}. This new mesh also increases the connection length by 20\% (see Table~\ref{tab1}), which is negligible compared to the difference in connection length by a factor 2 between the PT and NT shaping. Similarly, to understand the impact of radial transport, UEDGE simulations in the NT shaping are performed using the transport coefficients derived from the PT shaping. Finally, to test the effect of smaller $|B|$ in the NT divertor, a factor of $0.7$ is applied to the $\bf{E\times B}/B^2$ drift velocity to allow reduced such flows in the NT divertor. The factor of $0.7$ is obtained by the ratio of $R_\mathrm{X,PT}/R_\mathrm{X,NT}\approx 0.7$, where $R_\mathrm{X,PT}$ is the major radius of the X-point in the PT case and $R_\mathrm{X,NT}$ its value in the NT case. To evaluate impact of the differences between the NT Ohmic and PT Ohmic cases, just mentioned, 4 new simulations are performed.  These are as follows: (1), the NT parameters but using the extended mesh in Fig.~\ref{NTPT_grids}. (2), the initial NT mesh, but with $D_\perp$ and $\chi_\perp$ from the PT case. (3), combining the effects in 1 and 2. (4), the initial NT mesh and anomalous transport coefficients, but with the $\bf{E\times B}/B^2$ velocities reduced as described in the previous paragraph.
\begin{table}
\caption{\label{tab1} Parallel (Lpar) and poloidal (Lpol) Lengths from the outer midplane (OMP) to the outer target and the X-point to the outer target (outer leg) for three geometries in Fig.~\ref{NTPT_grids}, {\it i.e.}, NT, PT and NT with the extended leg.}
\footnotesize
\begin{tabular}{@{}cccc}
\br
    & NT & PT & NT (extended leg)\\
    & Lpar/Lpol (m) & Lpar/Lpol (m) & Lpar/Lpol (m)\\
\mr
 OMP to Target    & 9.56/1.06 & 31.9/1.78 & 11.5/1.15\\
 X-point to Target & 2.29/0.057 & 12.4/0.193 & 4.26/0.19\\
\br
\end{tabular}\\
\end{table}
\normalsize
Density ramps for the four cases by increasing the boundary density at the core-edge interface are performed. The results of these variation on $I_\mathrm{sat}$ and peak $T_e$ on the outer target plate as functions of $n_\mathrm{e,sep}$ are shown in Fig.~\ref{Fig:14} together with the initial NT and PT scans. It can be inferred that although the smaller $|B|$ in the divertor of the NT case enhances the $E\times B$ drift and strengthens the in-out divertor asymmetry in attached states~\cite{Scotti25}, its effect on divertor detachment remains limited. From the simulation results of the 'NT-enhanced-coef' case the enhanced transport coefficients reduce the $n_\mathrm{e,sep}$ required for detachment onset in the NT shaping by $\sim 23\%$. As a comparison, the extended divertor leg is able to reduce it by $\sim 33\%$. The combination of the two effects is responsible for $\sim 75\%$ of the difference between the $n_\mathrm{e,sep}$ required to achieve detached outer divertor in the NT and PT shaping. The remaining difference is likely attributable to the intrinsically shorter connection length in the NT shaping, whose effects are not covered in this study.

\section{Summary}\label{summary}

UEDGE simulations reveal that deep detachment of the outer divertor is not achievable in negative triangularity plasmas with reverse $B_T$, consistent with experimental observations of only shallow detachment. Both simulations and DIII-D measurements show that, although the onset of detachment occurs at a lower line-averaged density in reverse $B_T$ compared to forward $B_T$, the outer divertor fails to develop into a deeply detached state. In the reverse $B_T$ configuration, the $E \times B$ drift drives plasma toward the outer far-SOL, preventing density accumulation at the OSP. As a result, the UEDGE simulations show that with density scan, no further drop of $T_e$ at the OSP steady-state solutions when it reaches $\sim 2\mathrm{eV}$, followed by a transition to an unstable branch where the inner divertor undergoes a radiation-driven thermal collapse. This behavior provides a physics-based explanation for the experimentally observed saturation of $I_\mathrm{sat}$ at the OSP in reverse $B_T$ NT discharges despite continued gas puffing.

UEDGE simulations show that even in attached conditions, HFS radiation front already moves above the X-point in the forward $B_T$ configuration, which provides a physics explanation for the continuous confinement degradation observed experimentally. For forward $B_T$, $E \times B$ flows transport plasma from the outer to the inner divertor through the private flux region, the HFS radiation front moves quickly above the X-point even when the outer target remains in an attached state with $T_{e,\mathrm{OSP}} \sim 10$ eV. This causes a cooling of $T_e$ above the X-point inside the separatrix, which likely contributes to the experimentally observed degradation of core performance as density is increased prior to detachment onset. Despite this degradation, a series of stable, smoothly varying steady-state solutions are found with UEDGE throughout the density ramp in forward $B_T$, including stable solution of deeply detached outer divertor as HFS radiation extends over the X-point.

A comparative analysis between PT and NT Ohmic discharges, both experimentally and in UEDGE, shows that PT plasmas achieve detachment onset at $n_{\mathrm{e,sep}} \approx 1.4 \times 10^{19},\mathrm{m}^{-3}$, whereas NT plasmas fail to detach even at $n_{\mathrm{e,sep}} \approx 2.1 \times 10^{19},\mathrm{m}^{-3}$. The simulations reproduce this difference and indicate that the intrinsic difficulty of achieving a detached outer divertor in negative triangularity configurations is primarily due to reduced plasma transport and shortened divertor leg lengths. A further contribution from the inherently shorter connection length is also highly plausible, although this effect is not directly investigated in the present study.

\section{Acknowledgments}
This work was performed under the auspices of the U.S. Department of Energy by Lawrence Livermore National Laboratory under Contract DE-AC52-07NA27344 (LLNL-JRNL-2014421), and supported by the U.S. Department of Energy, Office of Fusion Energy Sciences, using the DIII-D National Fusion Facility, a DOE Office of Science user facility, under Award(s) DE-FC02-04ER54698.

\noindent Disclaimer: This report was prepared as an account of work sponsored by an agency of the United States Government. Neither the United States Government nor any agency thereof, nor any of their employees, makes any warranty, express or implied, or assumes any legal liability or responsibility for the accuracy, completeness, or usefulness of any information, apparatus, product, or process disclosed, or represents that its use would not infringe privately owned rights. Reference herein to any specific commercial product, process, or service by trade name, trademark, manufacturer, or otherwise does not necessarily constitute or imply its endorsement, recommendation, or favoring by the United States Government or any agency thereof. The views and opinions of authors expressed herein do not necessarily state or reflect those of the United States Government or any agency thereof.

\end{document}